\documentclass[twocolumn,english,aps,prb]{revtex4}
\usepackage[T1]{fontenc}
\usepackage[latin9]{inputenc}
\usepackage{graphicx}
\usepackage{amssymb}

\makeatletter

\newcommand{\noun}[1]{\textsc{#1}}

\makeatletter

\makeatletter

\makeatletter

\makeatletter

\usepackage{dcolumn}

\usepackage{bm}

\makeatother

\usepackage{babel}

\makeatother

\usepackage{babel}

\makeatother

\usepackage{babel}

\makeatother

\usepackage{babel}

\makeatother

\usepackage{babel}

\begin{document}

\title{Coexisting \emph{on-} and \emph{off-center} Yb$^{3+}$ sites in Ce$_{1-x}$Yb$_{x}$Fe$_{4}$P$_{12}$
skutterudites}

\author{F. A. Garcia$^{1}$, D. J. Garcia$^{2}$, M. A. Avila$^{1}$, J.
M. Vargas$^{1}$, P. G. Pagliuso$^{1}$, C. Rettori$^{1}$, M. C.
G. Passeggi$^{3}$, S. B. Oseroff$^{4}$, P. Schlottmann$^{5}$, B.
Alascio$^{2}$, and Z. Fisk$^{6}$}

\affiliation{$^{1}$Instituto de Física {}``Gleb Wataghin\textquotedbl{}, UNICAMP,
Campinas-SP, 13083-970, Brazil.\\
 $^{2}$Consejo Nacional de Investigaciones Científicas y Técnicas
(CONICET) and Centro Atómico Bariloche, S.C. de Bariloche, Río Negro,
Argentina.\\
 $^{3}$INTEC (CONICET and UNL), S3000GLN, Santa Fe, Argentina.\\
 $^{4}$San Diego State University, San Diego, California 92182, USA.\\
 $^{5}$ Department of Physics, Florida State University, Tallahassee,
Florida 32306, USA\\
 $^{6}$University of California, Irvine, CA, 92697-4573, USA.}

\date{\today}

\begin{abstract}
Electron Spin Resonance (ESR) measurements performed on the filled
skutterudite system Ce$_{1-x}$Yb$_{x}$Fe$_{4}$P$_{12}$ ($x\lesssim0.003$)
unequivocally reveal the coexistence of two Yb$^{3+}$ resonances,
associated with sites of considerably different occupations and temperature
behaviors. Detailed analysis of the ESR data suggests a scenario where
the fraction of oversized (Fe$_{2}$P$_{3}$)$_{4}$ cages that host
Yb ions are filled with a low occupation of \emph{on-center} Yb$^{3+}$
sites and a highly occupied $T$-dependent distribution of \emph{off-center}
Yb$^{3+}$ sites. Analysis of the $^{171}$Yb$^{3+}$(I=1/2) isotope
hyperfine splittings reveal that these two sites are associated with
a low ($\sim1$ GHz) and a high ($\gtrsim15$ GHz) rattling frequency,
respectively. Our findings introduce Yb$^{3+}$ in T$_{h}$ symmetry
systems and uses the Yb$^{3+}$ ESR as a sensitive microscopic probe
to investigate the Yb$^{3+}$ ions dynamics. 
\end{abstract}
\maketitle

\section{Introduction}

The dynamics of guest or filler ions vibrating loosely inside oversized
host cages has been a topic of current focus in condensed matter physics.
The anomalous behaviors of these so-called \emph{rattler} ions raise
interest both from the fundamental understanding of the unusual potential
wells they are subjected to (and consequent anharmonicities in their
vibrational motions) as well as from the implications of such \emph{rattling}
on the dampening of thermal transport in the material, which invites
application perspectives in the field of thermoelectrics.\cite{Snyder}
Thermoelectric materials, which can convert heat into electricity,
are of great interest for energy sustainability and energy harvesting
(transformation of waste heat into useful electricity). The main obstacle
is the low thermoelectric efficiency of materials for heat to electricity
conversion, which is quantified by the thermoelectric figure of merit,
$ZT$. The high $ZT$ value is the result of the high Seebeck coefficient
and the low thermal conductivity.\cite{Nolas} Among the best-known
cage systems displaying such characteristics are the filled skutterudite
compounds RT$_{4}$X$_{12}$, where R is a rare earth or actinide,
T is a transition metal (Fe, Ru, Os) and X is a pnictogen (P, As,
Sb). Besides exhibiting a rich variety of ground states and promising
thermoeletricity,\cite{Kobe2007} the question of whether the R ions
in these compounds are sited \emph{on-} and/or \emph{off-center} in
the oversized rigid (T$_{2}$X$_{3}$)$_{4}$-cages is a matter of
intense debate. \cite{Yanagisawa,Goto} There is also controversy
over the extent to which the weakly bounded R ions can be regarded
as independent Einstein oscillators, and how effectively they contribute
to a phonon-glass type of heat conduction.\cite{Lee,Vining,Koza}
In this work we take advantage of a uniquely favorable conjunction
between the chemical and structural characteristics of skutterudites
and their effect on the Electron Spin Resonance (ESR) of the $J=7/2$
multiplet of Yb$^{3+}$, to probe this ion's dynamical behavior within
oversized cages of the Ce$_{1-x}$Yb$_{x}$Fe$_{4}$P$_{12}$ system
($x\lesssim0.003$).

Skutterudites crystallize in the cubic LaFe$_{4}$P$_{12}$ structure
with space group $Im3$.\cite{Jeitschko} Each R ion is surrounded
by eight transition metal ions forming a cube, and twelve pnictogen
ions that form a slightly deformed icosahedron. Our work lies in the
fact that the local point symmetry for the R ions is T$_{h}$, which
lacks two symmetry operations ($C_{4}$ and $C_{2}^{\prime}$ rotations)\cite{Inui}
when compared to common cubic structures. Thus, the electric crystal
field (CF) Hamiltonian ($H_{CF}$) allows for an additional sixth
order term with an extra crystal field parameter (CFP), $B_{6}^{t}$.\cite{Takegahara,Bleaney}
This systems present a complex magnetic behavior, and it is essential
to know their CF level schemes for its complete description.\cite{references,Gorem}

ESR is a powerful microscopic tool to provide information about CF
effects, site symmetries, valencies of the paramagnetic ions, $g$-values,
fine and hyperfine parameters.\cite{Bleaney} The ESR of excited states
may be also observable, then, by measuring a R-ion ESR at different
frequencies and temperatures, one may obtain CF ground states and,
in some cases, the full set of CFP's that determine the overall splitting
of a R-ion $J$-multiplet ground state.\cite{Garcia} Previous works
on Ce$_{1-x}$R$_{x}$Fe$_{4}$P$_{12}$ for R = Nd, Dy, Er, Yb; $(x\lesssim0.005$)
succeeded in explaining the low-$T$ ESR results using such an expanded
$H_{CF}$, and the full set of CFP's could be determined.\cite{Garcia}
However, we now found that in the case of R = Yb, as $T$ increases,
a second Yb$^{3+}$ resonance emerges from the low-$T$ spectra, corresponding
to a distinct site, coexisting with the first one. The presence of
the new term in the $H_{CF}$ has proven essential to explain the
appearance of this second Yb$^{3+}$ resonance. The sensitivity of
Yb$^{3+}$ $4f$-electrons to this type of CF environment make it
a rare and useful probing ion to help the understanding of the potential
well responsible for its motion.

\section{Experimental}

Single crystals of Ce$_{1-x}$Yb$_{x}$Fe$_{4}$P$_{12}$ ($x\lesssim0.003$)
were grown in Sn-flux as described in Ref. \onlinecite{Meisner}.
The cubic structure ($Im3$) and phase purity were checked by x-ray
powder diffraction. The Yb concentrations were determined from the
$H$ and $T$-dependence of the magnetization, $M(H,T)$, measured
in a SQUID $dc$-magnetometer. The ESR experiments used crystals of
$\sim$$2$x$2$x$2$ mm$^{3}$ of naturally grown crystallographic
faces, as well as crystals crushed into fine powder. The ESR spectra
were taken in Bruker X ($9.48$ GHz) and Q ($34.4$ GHz) band spectrometers
using appropriate resonators coupled to a $T$-controller of a helium
gas flux system for $4.2\lesssim T\lesssim45$ K. The ESR spectra
of the $^{170}$Yb$^{3+}$(I=0) isotope showed the superposition of
a \emph{narrow line} and a slightly shifted \emph{broad line}. For
single crystals and powdered samples the spectra were, respectively,
fitted by the superposition of two dysonian (metallic lineshape) and
two lorentzian resonances with adjustable resonance fields ($H_{0}$),
linewidths ($\Delta H$), A/B ratios and amplitudes.\cite{Feher}


\section{Experimental results}

\begin{figure}[thpb]
 \includegraphics[width=68mm]{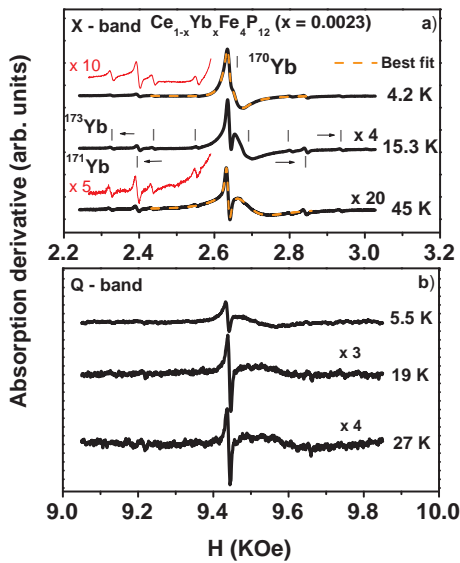}

\caption{(color online) $T$-evolution ($4.2\lesssim T\lesssim40$ K) of the
normalized Yb$^{3+}$ ESR spectra in a Ce$_{1-x}$Yb$_{x}$Fe$_{4}$P$_{12}$
($x\simeq0.0023$) single crystal: a) X-band and b) Q-band. The enhanced
low field spectra shows the absence of hyperfine lines for the \emph{broad
line}. \label{Fig1}}

\end{figure}

\begin{figure}[thpb]
\includegraphics[width=72mm]{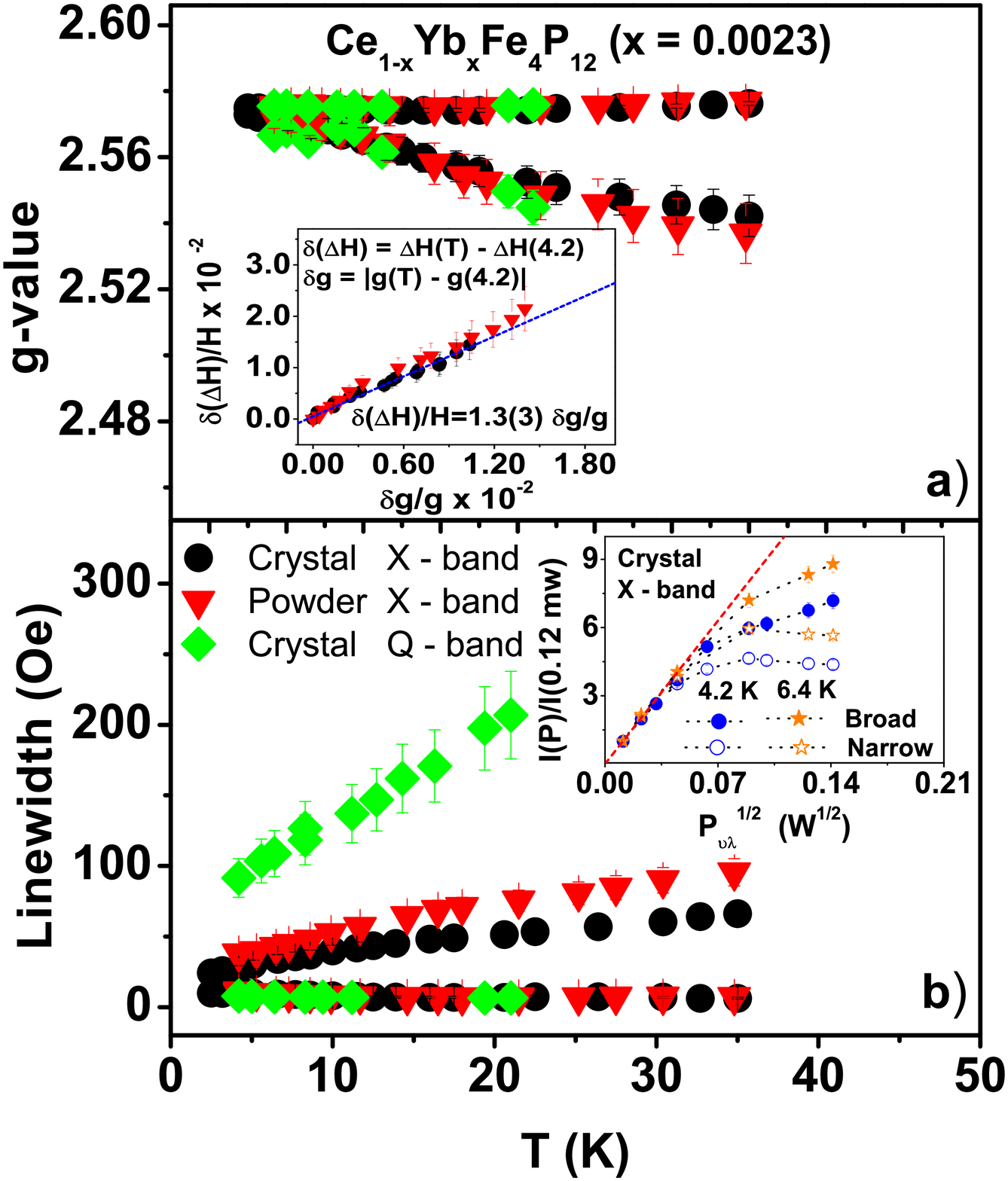}

\caption{(color online) X and Q-bands low $T$-evolution of: a) $g$-value
and b) $\Delta H$ for the \emph{narrow} and \emph{broad lines} of
Fig. 1 and also for a powdered crystal. At X-band, Inset a) shows
the correlation between $\delta$$(\Delta H)/H$ and $\delta g/g$,
and Inset b) the microwave power dependence of the ESR intensity.
\label{Fig2}}

\end{figure}

Figures 1a and 1b show, respectively, selected X- and Q-band ESR spectra
for the Kramers doublet ground state (KDGS) of the $^{170}$Yb$^{3+}$(I=0)
isotope in a Ce$_{1-x}$Yb$_{x}$Fe$_{4}$P$_{12}$ ($x\cong0.0023$)
single crystal. As $T$ increases the nearly single line observed
at low-$T$ for the $^{170}$Yb$^{3+}$ isotope evolves into two lines,
a \emph{narrow} and a \emph{broad} one. At low-$T$ the measured $g$-values
for the \emph{narrow} and \emph{broad lines} are essentially the same,
$g\simeq$ 2.57, different from the $g$-values of $2.666$ and $3.428$
expected for $\Gamma_{6}$ and $\Gamma_{7}$ doublets, respectively.\cite{Bleaney}
At X-band the \emph{narrow line} displays the full hyperfine spectra
for the Yb isotopes $^{170}$Yb$^{3+}$(I=0), $^{171}$Yb$^{3+}$(I=1/2)
and $^{173}$Yb$^{3+}$(I=5/2), confirming that the observed spectra
are associated to Yb$^{3+}$ ions. From the hyperfine splittings the
corresponding hyperfine constants $^{171}A$ = 440(10) Oe and $^{173}A$
= 120(3) Oe were obtained. These values are $\sim20\%$ smaller than
the hyperfine constants of Yb$^{3+}$ in a KDGS of any system with
$O_{h}$ cubic point symmetry.\cite{Bleaney,Chock} The hyperfine
lines corresponding to the \emph{broad line} of the $^{171}$Yb$^{3+}$
isotope were not observed.

\begin{figure}[hptb]
\includegraphics[width=72mm]{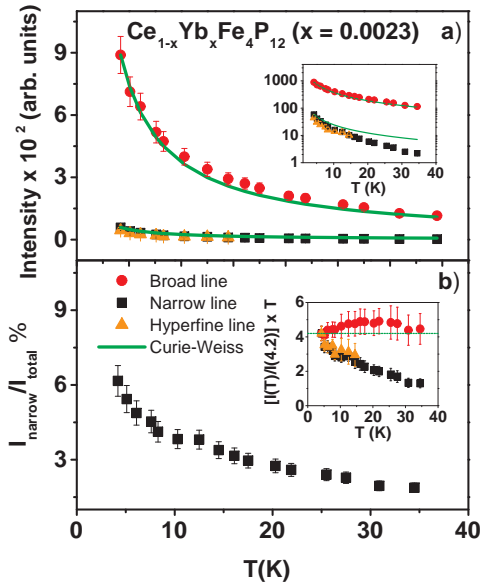}

\caption{a) $T$-dependence of the ESR intensity for the two resonances of
the $^{170}$Yb$^{3+}$(I=0) isotope. The intensity of one of the
$^{171}$Yb$^{3+}$(I=1/2) hyperfine line, normalized by the natural
abundance, is also shown. The inset presents the same data in log
scale. b) $T$-dependence of the relative intensity for the \emph{narrow
line}. The inset displays the data of Fig. 3a ($\times T$) normalized
at 4.2 K. The green lines show the C-W behavior.}

\label{Fig3} 
\end{figure}

Figures 2a and 2b show, respectively, the $T$-evolution ($4.2\lesssim T\lesssim40$
K) of the $g$-values and linewidths, $\Delta H$, for the \emph{narrow}
and \emph{broad lines} of Fig. 1 and also for a powdered sample at
X-band. The following features are noteworthy: \emph{a}) for the sites
corresponding to the \emph{narrow line} the $g$-value and $\Delta H$
are frequency- and $T$-independent; \emph{b}) for the \emph{broad
line} sites the $g$-value and $\Delta H$ are $T$-dependent, and
only $\Delta H$ is frequency dependent; \emph{c}) the \emph{broad
line} of the powdered sample is broader than that of the crystal,
while the \emph{narrow line} is about the same; \emph{d}) angular
dependent ESR experiments found these resonances to be isotropic;
\emph{e}) for the various samples the relative change of the \emph{broad
line} linewidth, $\delta$$(\Delta H)/H$, scales at all-$T$ with
the relative change of its $g$-value, $\delta g/g$ ($\delta(\Delta H)/H\simeq$
1.3(3) $\delta g/g$, see inset of Fig. 2a); \emph{f}) saturation
ESR intensity measurements show that at 4.2 K and $\sim$ 10 mW the
\emph{broad} and \emph{narrow lines} present, respectively, a $\sim30\%$
and $\sim50\%$ saturation (see inset of Fig. 2b).

All the experimental features given in Figs. 1 and 2 were confirmed
in crystals from different batches with comparable Yb concentrations.
They lead us to conclude that: \emph{i)} the \emph{narrow} and the
\emph{broad lines} are, respectively, homogeneous and inhomogeneous
resonances; \emph{ii)} the origin of the inhomogeneity is a distribution
of $g$-values of the order of the change in the $g$-value; \emph{iii)}
the $T$-independent $\Delta H$ for the \emph{narrow line} indicates
that there is no Yb$^{3+}$ spin-lattice relaxation via exchange interaction
with the conduction-electrons; \cite{Korringa,Rettori} \emph{iv)}
the saturation of the spectra at low-$T$ suggests slow spin-lattice
relaxation involving lattice phonons via spin-orbit coupling;\cite{Bleaney}
and \emph{v)} at low-$T$ the Yb$^{3+}$ ions behave as an \emph{adiabatic}
spin system allowing the formation of Einstein oscillators inside
the (Fe$_{2}$P$_{3}$)$_{4}$-cages.

Figure 3a displays the $T$-dependence of the unsaturated ($\sim$2
mW) ESR intensity for the \emph{broad} and \emph{narrow lines} of
the $^{170}$Yb$^{3+}$(I=0) isotope of Fig. 1a. The intensity of
one of the $^{171}$Yb$^{3+}$(I=1/2) isotope hyperfine lines, normalized
by its natural abundance, is also shown. The inset shows the data
in a log scale. From their relative intensities the \emph{broad} and
\emph{narrow lines} correspond, respectively, to $\sim95\%$ and $\sim5\%$
of the Yb$^{3+}$ ions filling cages. Figure 3b presents the $T$-dependence
of the relative population for the low occupied sites (\emph{narrow
line}). The large observed drop strongly suggests that, as $T$-increases,
the low populated Yb$^{3+}$ sites migrate, in a reversible way, to
the highly populated Yb$^{3+}$ ones. The inset of Fig. 3b shows for
the \emph{broad}, \emph{narrow} and hyperfine lines the $T$-dependence
of their intensities ($\times T$) normalized at $T\simeq$ 4.2 K.
This data reveals that the \emph{broad line} practically follows a
Curie-Weiss (C-W) law, while the \emph{narrow line} surprisingly drops
faster than a C-W behavior, given further support to the sites migration
idea. The C-W behavior is another indication that the $^{170}$Yb$^{3+}$
ions carry localized magnetic moment and that the resonances arise
from a CF KDGS.



\section{Analysis and discussion}

In order to analyze our ESR data in Ref. 15 we used the expanded Hamiltonian,
$H_{CFZ}$:

\begin{eqnarray}
H_{CFZ} & = & W\left\{ (1-\vert\mathrm{y}\vert)\left[\mathrm{x}\frac{O_{4}^{c}}{F_{4}^{0}}+(1-\vert\mathrm{x}\vert)\frac{O_{6}^{c}}{F_{6}^{0}}\right]+\mathrm{y}\frac{O_{6}^{t}}{F_{6}^{2}}\right\} \nonumber \\
 &  & +g_{J}\mu_{B}{\bf H}\cdot{\bf J},\label{HCFZ}\end{eqnarray}

where a magnetic moment \noun{J} with a Landé \emph{g}-factor $g_{J}$
is considered. The CF includes the usual cubic O$_{h}$ \cite{Lea}
terms parametrized by the x variable that measures the relative weight
of the 4th and 6th order terms and also takes into consideration the
new term $O_{6}^{t}$. The relative weight y linearly interpolates
between the O$_{h}$ cubic terms for y$=0$ and the $O_{6}^{t}$ term
for y$=1$. This (x,y) parametrization allows the entire range of
the CFP's to be accounted for within the finite intervals $-1\leq$x$\leq1$
and |y|$<1$ and the results do not depend on the sign of y. By diagonalizing
$H_{CFZ}$ one obtains, as a function of x and y, the CF wave functions
and energies for each of the R in units of W. From the ground state
wave function the low field $g$-value can be calculated \cite{Garcia}.



A combined analysis of the ESR data for Er$^{3+}$, Dy$^{3+}$ and
Yb$^{3+}$ impurities in Ce$_{1-x}$R$_{x}$Fe$_{4}$P$_{12}$ allowed
us to pinpoint the exact (x=0.523,y=0.082) values corresponding to
the Yb$^{3+}$ \emph{narrow line} observed at $T\simeq$ 4.2 K and
$g\simeq2.575$.\cite{Garcia} However, as $T$ increases, a second
Yb$^{3+}$ \emph{broad line} emerges from the low-$T$ \emph{narrow
line} (Fig. 1) and its $g$-value decreases, reaching $g$ = 2.54(1)
at our highest-$T$ ($\simeq$ 45 K). Therefore, these two resonances
should be associated to two coexisting Yb$^{3+}$ sites with different
peculiarities.

In these compounds the R-ions are known to rattle at frequencies of
$\sim10^{3}$ GHz \cite{Lee,Vining,Koza} which are low compared to
the cage ion vibrations, but still much higher than the ESR frequencies
($\sim$10-30 GHz). Thus, we argue that the reduced hyperfine constant
for the homogeneous \emph{narrow line} spectra results from a motional
narrowing mechanism \cite{Farach} of \emph{on-center} Yb$^{3+}$
ions rattling in the rigid oversized ($\phi\simeq5$ \AA) (Fe$_{2}$P$_{3}$)$_{4}$-cages.\cite{Jeitschko}
In the extreme motional narrowing regime\cite{Anderson} a rattling
frequency of $\sim1$ GHz will reduce in $\sim20\%$ the hyperfine
constant. Moreover, the hyperfine structure in the inhomogeneous \emph{broad
line} spectra was not observed, suggesting that a distribution of
Yb$^{3+}$ ions are rattling at higher frequencies and producing an
even larger reduction of the hyperfine constant. Again, in the extreme
motional narrowing regime, a rattling frequency $\gtrsim$15 GHz will
reduce in 90 to 95$\%$ the hyperfine splitting, and the ESR spectra
would look like the observed single \emph{broad line} of $\Delta H\simeq$
30-40 Oe. We should mention that the reported rattling amplitudes
are $\lesssim$ 0.1 \AA.\cite{Cao04} Hence, the \emph{broad line}
$T$-dependent shift and broadening is most likely the result of a
$T$-dependent distribution of Yb$^{3+}$ ions rattling at higher
frequencies inside the (Fe$_{2}$P$_{3}$)$_{4}$-cages. Thus, we
associate the homogeneous \emph{narrow line}, corresponding to the
low occupied sites, to \emph{on-center} ($g$ = 2.575) of $\simeq$1
GHz rattling Yb$^{3+}$ ions at (x=0.523,y=0.082), whereas the inhomogeneous
\emph{broad line}, corresponding to the highly occupied sites with
lower $g$-values, to a distribution of $\gtrsim$15 GHz rattling
Yb$^{3+}$ ions. Since this \emph{broad line} is an inhomogeneous
resonance (distribution of $g$-values) and the rattling frequency
is of the order of or higher than the ESR frequency, the rattling
Yb$^{3+}$ ions responsible for these spectra should be spending more
time at \emph{off-center} positions in the over-size cage.

Since no emerging second resonance was observed from the low-$T$
ESR spectra of Er$^{3+}$ and Dy$^{3+}$ ions diluted in Ce$_{1-x}$R$_{x}$Fe$_{4}$P$_{12}$,\cite{Garcia}
it is possible that for Yb$^{3+}$, with smaller ionic radius than
that of Er$^{3+}$ and Dy$^{3+}$, larger voided excursion space may
be available for the Yb$^{3+}$ ions inside the (Fe$_{2}$P$_{3}$)$_{4}$-cages
which may further favor the Yb$^{3+}$ to rattle. A $T$-dependent
distribution of CFPs, that in this T$_{h}$ symmetry allows for a
continuous change on $g_{eff}$\cite{F Garcia}, or even a distribution
of new 2$^{nd}$ order CFPs in $H_{CFZ}$ associated to the \emph{off-center}
Yb$^{3+}$ sites may be also a plausible reason for the observed $T$-dependence
of the inhomogeneous \emph{broad line}.


\section{Conclusions}

In summary, our ESR results have shown that the origin for the large
$g$-shift of the Yb$^{3+}$ KDGS, relative to that in O$_{h}$ symmetry
(y = 0), may be associated to the $B_{6}^{t}(O_{6}^{2}-O_{6}^{6})$
term in $H_{CFZ}$. Coexisting \emph{narrow} and \emph{broad} Yb$^{3+}$
resonances were observed and associated, respectively, to a low occupation
($\sim$5$\%$) of \emph{on-center} Yb$^{3+}$ rattling ions ($\sim$1
GHz) and to a highly occupied ($\sim$95$\%$) $T$-dependent distribution
of \emph{off-center} rattling Yb$^{3+}$ ions ($\gtrsim$15 GHz).
These assignments were based on: \emph{i)} the much higher expected
Yb$^{3+}$ rattling frequencies than the microwave frequency used
in the ESR experiments\cite{Lee,Vining,Koza} and; \emph{ii)} on the
observed reduction of the hyperfine constant for the \emph{on-center}
Yb$^{3+}$ ions and the absence of hyperfine structure in the spectra
of the \emph{off-center} Yb$^{3+}$ ions which were attributed to
motional narrowing effects.\cite{Farach,Anderson} Although our findings
relied on the Yb$^{3+}$ ESR results to witness the Yb$^{3+}$ rattling
mode, they suggest that the R ions in other skutterudites and clathrate
compounds may be also rattling in an analogous form as long as they
are inside an oversized cage. However, it may not be always observable
in an ESR experiment. We believe that the evidence for predominant
\emph{off-center} rattling Yb$^{3+}$ ions in these skuterudites is
a result that could justify the existence of Einstein oscillators
and help to understand the low thermal conductivity and the strongly
correlated phenomena exhibited by these type of materials. 

\section{Acknowledgments}

We thank FAPESP-SP and CNPq for financial support. PS is supported
by DOE grant No. DE-FG02-98ER45707.

\end{document}